\documentclass[11pt,a4paper]{article}
\pdfoutput=1
\usepackage{jheppub}
\usepackage[all]{xy} 
\begin{document}

\newcommand{\hphi}{\phi}
\newcommand{\bphi}{\varphi}
\newcommand{\hsigma}{{\sigma}}

\title{A Simple Holographic Model of a Charged Lattice}

\author{Francesco Aprile,}
\author{and Takaaki Ishii}
\affiliation{Crete Center for Theoretical Physics, Department of Physics, University of Crete, \\ PO Box 2208, 71003 Heraklion, Greece}
\emailAdd{aprile@physics.uoc.gr}
\emailAdd{ishii@physics.uoc.gr}

\abstract{We use holography to compute the conductivity in an inhomogeneous charged scalar background. We work in the probe limit of the four-dimensional Einstein-Maxwell theory coupled to a charged scalar. The background has zero charge density and is constructed by turning on a scalar source deformation with a striped profile. We solve for fluctuations by making use of a Fourier series expansion. This approach turns out to be useful for understanding which couplings become important in our inhomogeneous background. At zero temperature, the conductivity is computed analytically in a small amplitude expansion. At finite temperature, it is computed numerically by truncating the Fourier series to a relevant set of modes. In the real part of the conductivity along the direction of the stripe, we find a Drude-like peak and a delta function with a negative weight. These features are understood from the point of view of spectral weight transfer.   
}

\dedicated{{\rm CCTP-2014-11 \\ CCQCN-2014-30}}

\maketitle

\section{Introduction}
\label{sec:introduction}

The gauge/gravity duality offers an attractive framework for studying the dynamics of strongly interacting systems by using classical gravity computations. Of recent interest are the applications of the AdS/CFT correspondence to condensed matter physics. A notable step forward in this direction has been the construction of gravitational background in which the charge density is spatially modulated and translational invariance is explicitly broken \cite{Horowitz:2012ky, Horowitz:2012gs,Horowitz:2013jaa,Chesler:2013qla,Ling:2013nxa}. These systems are commonly denoted as ``holographic lattices".
Prior to this development, there have been studies of perturbative lattice solutions on top of a metallic phase
\cite{Maeda:2011pk,Hartnoll:2012rj,Liu:2012tr,Iizuka:2012dk,Ling:2013aya},
and in the context of striped superconductors 
\cite{Flauger:2010tv,Hutasoit:2011rd,Ganguli:2012up,Hutasoit:2012ib}.

In dissipative systems, the optical conductivity at small frequencies is well characterized by a Drude peak,
\begin{align}
\sigma(\omega)=\frac{K \tau}{ 1- i\omega \tau}\ ,
\end{align}
where $\tau$ is the relaxation time of the charge careers and $K$ is a constant. In these systems, the existence of $\tau$ signals the presence of low energy excitations at finite momentum which scatter the charge carriers. One way to introduce such excitations in the spectrum of a theory is to break translation invariance by using a spatially periodic source. The holographic lattices are the first concrete realization of this idea. 

There is also a broad class of effective models describing dissipative dynamics. An important example in this class is the massive gravity theory of \cite{Vegh:2013sk,Davison:2013txa,Blake:2013bqa,Davison:2013jba,Amoretti:2014zha}. In this theory, the graviton has an explicit mass term in the Lagrangian, and hence the translational invariance is absent. Other examples of the explicit breaking of translational invariance include Q-lattices \cite{Donos:2013eha,Donos:2014uba}, scalar sourced models \cite{Andrade:2013gsa,Gouteraux:2014hca}, and a model based on the Bianchi VII$_0$ symmetry \cite{Donos:2012js,Donos:2014oha}. It is possible to produce a Drude-like peak also in bottom-up models with specific couplings \cite{Ishii:2012hw}.

There is a direct connection between the massive gravity and the holographic lattice, which we would like to underline. What happens is the following: a subset of the perturbations relevant for calculating the conductivity in the holographic lattice reduces to that of massive gravity, in a gauge.  In this formalism, the mode which describes the vibration of the lattice is eaten up by the metric which, in turn, acquires a radially dependent effective mass. In analogy with condensed matter physics, this mode has been identified as a bulk phonon arising from the lattice. The virtue of the massive gravity is to highlight the most crucial ingredient that the lattice brings into this story: the coupling between the gauge field and the phonon. In the example of \cite{Blake:2013owa}, this coupling is through a specific non diagonal mass matrix.

As long as the gravitational background is spatially modulated, bulk phonons are present in the spectrum, and the optical conductivity shows the Drude peak. However, we would like to consider other situations in which dissipative effects are mediated by other types of fluctuations. For instance, in striped superconductors, fluctuations of the complex scalar may be also important to the dynamics of the system, and it would be interesting to understand how each fluctuation contributes to the conductivity. In particular, there may be different mechanisms, other than that of massive gravity, that determines Drude-like peak behavior in such systems.  Finally, fluctuations of the complex scalar may also affect the coherence of the superconducting state. Hence, it would be important to study them in lattice backgrounds.

In this paper, we consider a simple holographic model in which the background $U(1)$ charged scalar is spatially modulated with a stripe profile. We work in the probe limit and at zero charge density. We refer to this background as a ``charged lattice''. Our setup simplifies the analysis of fluctuations and allows us to focus on the effects coming from the coupling between the $U(1)$ current and the charged scalar field.  We compute the conductivity in the directions transverse and longitudinal to the stripe. We will see that, in the longitudinal direction, the conductivity is characterized by a Drude-like peak and by a delta function with a negative weight. We will discuss these two features in connection with spectral weight transfer.

This paper is organized as follows. In section~\ref{sec:model}, we introduce our model and the scalar background. In section~\ref{sec:generalCond}, we describe the Fourier decomposition that we use for calculating the conductivity. In section~\ref{sec:AdSCalculation}, we analytically compute the conductivity in the zero temperature AdS background by making a perturbative expansion in small stripe amplitude. In section~\ref{sec:numerical}, we numerically compute the conductivity at finite temperature by truncating the Fourier series to a relevant set of modes. Section~\ref{sec:conclusion} contains conclusions and discussion. Appendices contain some details of computations.

\section{The Model}
\label{sec:model}

We consider a four-dimensional theory with a Maxwell field $A_{\mu}$ and a charged complex scalar $\Phi$ in a curved background. The action is
\begin{align}
S = \int d^4x \sqrt{-g}
\left( -\frac{1}{4} F_{\mu\nu} F^{\mu\nu} - | \partial_{\mu} \Phi -  i A_{\mu} \Phi |^2 - m^2 \Phi^2 \right),
\label{swave_lagrangian}
\end{align}
where $F=dA$.  The mass of the scalar field is chosen to be $m^2=-2$.
We work in the probe limit where the backreaction of the matter fields to the metric is suppressed.
The background is the AdS-Schwarzschid black hole, whose metric can be given as
\begin{align}
ds^2 = \frac{1}{r^2} \left( -f(r) dt^2 + \frac{dr^2}{f(r)} + dx^2 + dy^2 \right), \quad f(r)=1-\frac{r^3}{r_{h}^3}.
\label{ads_sch_metric}
\end{align}
In our conventions, the AdS radius is set to one, and the gauge coupling is absorbed into a redefinition of the fields. The AdS boundary and the black hole horizon are located at $r=0$ and $r=r_h$, respectively. The Hawking temperature of the black hole is $T_{H}=3/(4\pi r_h)$. The action \eqref{swave_lagrangian} was studied as a model of s-wave holographic superconductivity \cite{Hartnoll:2008vx}.

The complex scalar $\Phi$ can be written in terms of a real scalar field $\hphi$ and a Stuckelberg field $\theta$ as $\Phi= {\hphi} \, e^{i\theta}/\sqrt{2}$. In these fields, the action becomes
\begin{align}
S = \int d^4x \sqrt{-g}
\left( -\frac{1}{4} F_{\mu\nu} F^{\mu\nu} - \frac{1}{2}(\partial_{\mu} {\hphi})^2-\frac{ {\hphi}^2}{2} (\partial_{\mu} \theta - A_{\mu} )^2 - m^2 {\hphi}^2 \right).
\end{align}
The equations of motion are
\begin{align}
\nabla_{\mu} F^{\mu\nu} - {\hphi}^2 \left( \partial^{\nu}\theta-A^{\nu} \right)&=0,
\nonumber\\
\partial_{\mu}\big(  {\hphi}^2 \sqrt{-g}\  (\partial^{\mu}\theta-A^{\mu}))&=0,
\nonumber\\
(\nabla_{\mu} \nabla^{\mu} - m^2 -(\partial^{\mu}\theta-A^{\mu})^2)  {\hphi} &=0.
\end{align}
In a background in which $A_{\mu}=0$ and $ {\hphi}$ is nonzero, $\theta$ is a constant.
We can choose $\theta=0$ by using the gauge transformation $A_{\mu} \to A_{\mu} +\partial_\mu \Lambda$, $\theta \to \theta+\Lambda$. The scalar field $\Phi$ is then real.

\subsection{The scalar profile}
\label{sec:stripebg}

We are interested in a scalar field configuration that is inhomogeneous in one spatial direction.
By considering the ansatz $A_{\mu}=0$ and $\Phi={\hphi}(r,x)/\sqrt{2}$, we derive the equation of motion of ${\hphi}$,
\begin{align}
\left( \partial_r^2 + \left( \frac{f'}{f} - \frac{2}{r} \right)\partial_r + \frac{1}{f} \partial_x^2 - \frac{m^2}{r^2 f} \right) {\hphi}(r,x) = 0, \label{phi_rx_eom}
\end{align}
where $f' \equiv \partial_r f$. 
At the AdS boundary, $r=0$, the field ${\hphi}(r,x)$ has the following series expansion,
\begin{align}
{\hphi}(r,x) = {\hphi}_1(x) r + {\hphi}_2(x) r^2 + \cdots \ .
\label{phi_bdry_series}
\end{align}
According to the AdS/CFT correspondence, $\Phi(r,x)$ is dual to a charged scalar operator $\mathcal{O}(x)$. When $m^2=-2$, both $\hphi_1(x)$ and $\phi_2(x)$ correspond to normalizable modes. It is known as the standard quantization to choose the dimension of $\mathcal{O}(x)$ as $\Delta=2$. In this case, $\hphi_1(x)$ and $\hphi_2(x)$ are interpreted as the source and the vacuum expectation value of $\mathcal{O}(x)$, respectively. Choosing $\Delta=1$ is called the alternative quantization, where the roles of $\hphi_1(x)$ and $\hphi_2(x)$ are interchanged compared with the case of the standard quantization. In this paper, we mainly consider the standard quantization. 

We introduce a periodic source deformation of the field theory by turning on
\begin{equation}
\hphi_1(x)=V \cos (Q x).
\end{equation}
Since eq.~\eqref{phi_rx_eom} is linear, we can take the bulk field ${\hphi}(r,x)$ to have the form,\footnote{In general, the background scalar may be given by
\begin{align}
{\hphi}(r,x) = \sum_{n=0}^{\infty} \left( \bphi_\mathrm{even}^{(n)} (r) \cos (n Q x) + \bphi_\mathrm{odd}^{(n)} (r) \sin (n Q x) \right).
\label{phibg_general_ansatz}
\end{align}
Since eq.~\eqref{phi_rx_eom} is linear, the modes $\bphi_\mathrm{even}^{(n)}(r)$ or $\bphi_\mathrm{odd}^{(n)}(r)$ do not couple each other.}
\begin{align}
{\hphi}(r,x) = \bphi(r) \cos (Q x),
\label{phibg_cos_ansatz}
\end{align}
where the parameter $Q$ is the momentum associated to the striped profile \eqref{phibg_cos_ansatz}. 
With this ansatz, eq.~\eqref{phi_rx_eom} becomes
\begin{align}
\left( \partial_r^2 + \left( \frac{f'}{f} - \frac{2}{r} \right)\partial_r - \frac{m^2 + r^2 Q^2}{r^2 f} \right) \bphi(r) = 0, 
\label{phi_r_eom}
\end{align}
The momentum $Q$ gives a positive contribution to the bulk mass of $\bphi$, which vanishes at the boundary. At $r=0$, the series expansion is $\bphi(r)=\bphi_1 r+ \bphi_2 r^2 +\cdots$. We deduce that $\bphi_1=V$, whereas
 $\bphi_2$ corresponds to the amplitude of $\langle \mathcal{O} \rangle$.

\begin{figure}[t]
\centering
\includegraphics[width=6cm]{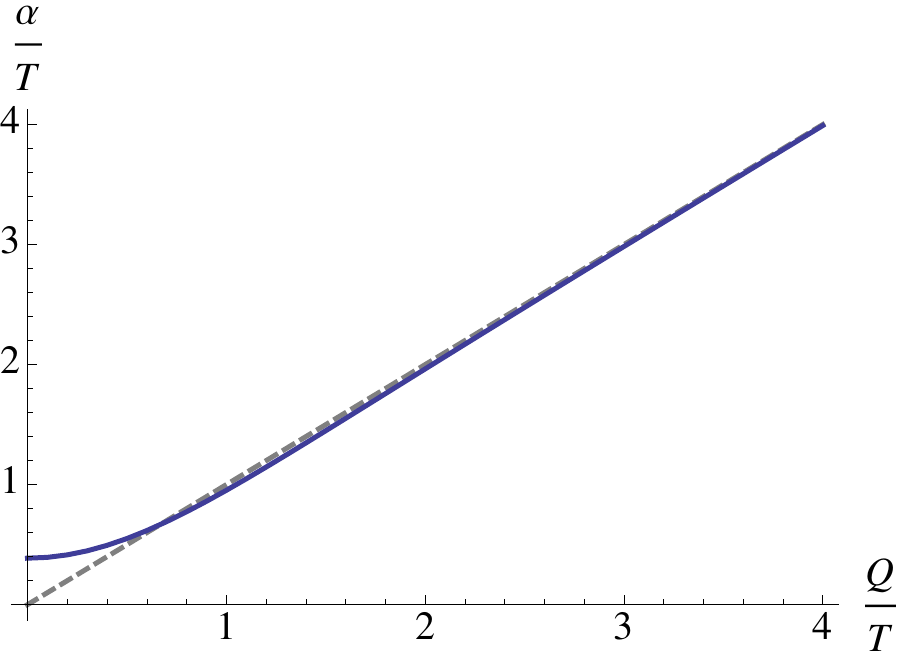}
\caption{Dependence of $\alpha/T$ on $Q/T$. The gray dashed line is $\alpha=Q$.}
\label{fig:QalphaPlot}
\end{figure}

The field $\bphi(r)$ is obtained by solving eq.~\eqref{phi_r_eom} numerically. Regularity at the horizon implies
\begin{align}
\bphi'(r_h)=-\frac{m^2 + r_h^2 Q^2}{3}\bphi(r_h).
\end{align}
In general, solutions are parametrized by two dimensionless parameters built out of $V$, $Q$, and the temperature $T$.\footnote{We normalize our $T$ as $T=4\pi T_H/3$.} Because we work in the Schwarzschid black hole background and the eq.~\eqref{phi_r_eom} is linear, the ratio $\alpha\equiv -\bphi_2 /\bphi_1$ has the property that $\alpha/T$ does only depend on $Q/T$.
In figure~\ref{fig:QalphaPlot}, we plot $\alpha/T$ as a function of $Q/T$. In the large $Q$ limit, we find that  $\alpha(Q)=Q$. At $Q=0$, we obtain $\alpha(0)/T = 0.387$.\footnote{
When $Q=0$, eq.~(\ref{phi_r_eom}) can be analytically solved, and $\alpha$ can be computed as (See e.g.~\cite{Faulkner:2010gj}.)
\begin{align}
\frac{\alpha}{T}=\frac{\Gamma(2/3)^3}{\Gamma(4/3) \, \Gamma(1/3)^2} \approx 0.387.
\end{align}
}

By construction, the average value of the source ${\hphi}_1(x)$ vanishes along the $x$ direction. The homogeneous mode of the scalar field is therefore not excited. 
On the other hand, the inhomogeneous mode ${\hphi}(r,x)$ gets a contribution to its bulk mass from the momentum $Q$, and its amplitude mode $\varphi(x)$ tends to decrease as the horizon is approached. 
Given the property that ${\hphi}_1(x)$ has zero average value along the $x$ direction, we may interpret our scalar profile as a special kind of charged impurity background of the boundary theory. We refer to this background as a ``charged lattice''.

\section{Optical Conductivity}
\label{sec:generalCond}
The optical conductivity is defined by the general expression,
\begin{align}\label{def_cond_1}
 {J}^i(\vec{x}) =\ \sigma^{ij}(\vec{x}) {E}_{j},
\end{align}
where $\vec{E}$ is an applied electric field and $\vec{J}$ is the induced current. In this section, we explain how to calculate the conductivity in our holographic setup.

According to the holographic dictionary, the electric field $\vec{E}$ and the current $\vec{J}$ of the dual field theory are given by
\begin{align}\label{def_cond_2}
{J}^a \equiv \lim_{r\rightarrow 0} \sqrt{-g}f^{a r}\ ,  \qquad {E}_j \equiv\lim_{r\rightarrow 0} f_{jt}\ , 
\end{align} 
where $f_{\mu\nu}$ is the bulk field strength of gauge field fluctuation $\delta A_{\mu}$. To avoid confusion, the index $a$ refers to the $(2+1)$-dimensional spacetime, $a=t,x,y$, whereas the index $i$ labels the two spatial directions $i = x,y$. The definition (\ref{def_cond_2}) is manifestly gauge invariant. We choose the gauge $\delta A_r=0$. The current becomes
\begin{align}
{J}^{a} =-\lim_{r\rightarrow 0} \sqrt{-g} g^{rr}g^{ab}\partial_r \delta A_{b}\ ,
\end{align}
and the conductivity can be obtained from the definition (\ref{def_cond_1}). In our calculation, we will consider a homogeneous electric field with a time dependence of the form $e^{-i\omega t}$, and study in detail the average part of the conductivity, denoted as ${\hsigma}(\omega)$.

On top of our background, $A_{\mu}=0$ and ${\hphi}$ given in \eqref{phibg_cos_ansatz}, generic perturbations have the form,
\begin{align}
 A_{\mu} \to \delta A_{\mu},\qquad \Phi \to {\hphi}/\sqrt{2}+ (\delta\eta+ i\delta\psi)/\sqrt{2} \ .
\end{align}
Because of the inhomogeneity of the scalar field, fluctuations of the gauge field in $x$- and $y$-directions behave differently. The consistent sets of perturbations are given as follows:
\begin{itemize}
\item
When an electric field is applied in the direction transverse to the stripe,
\begin{align}
\delta A_y=a_y(r,x) e^{-i\omega t}\ .
\end{align}
\item
When an electric field is applied in the direction longitudinal to the stripe,\footnote{A related work on lattice effects was given in \cite{Iizuka:2012dk}. In this paper, we explicitly take into account the charged scalar fluctuation.}
\begin{align}
\delta A_x= a_x(r,x) e^{-i \omega t}, \quad \delta A_t=a_t(r,x) e^{-i \omega t}, \quad \delta\psi=\psi(r,x)e^{-i \omega t}\ .
\label{longitudinal_fluctuation_ansatz}
\end{align}
\end{itemize}
We refer to these two sets of perturbations as transverse and longitudinal channels, respectively. 
The longitudinal channel involves the dynamics of the scalar field fluctuation. The fluctuation of the real part of the scalar field, $\delta\eta$, decouples from the longitudinal channel because the background gauge field is absent. 

Thinking in more general terms, we may consider a bulk theory with a gauge group $G$ and a scalar field $\Phi$ transforming in a non trivial representation of $G$. For example, if the gauge group is $SU(2)$, the dual field theory may be regarded as the low energy effective description of a spin variable  and the scalar field may be interpreted as the order parameter for  antiferromagnetism or it may describe a spin density wave phase \cite{Iqbal:2010eh}. Then, certain fluctuations of such a order parameter corresponds to bulk pions. In this sense, our $\delta\psi$ produces a vibration of the striped background in the $U(1)$ manifold and plays a different role with respect to the bulk phonon in massive gravity.  

In order to describe the fluctuations of our background, we could use gauge invariant variables, $B_{\mu}=A_{\mu}-\partial_{\mu}\theta$, instead of working in the $\delta A_r=0$ gauge. In this case, the effective description of the system is regarded as a theory of a massive $U(1)$ gauge field with a radial dependent mass and a dynamical $B_r$ which takes the place of the scalar field fluctuation $\delta\psi$. We would like to stress that taking into account the dynamical $B_r$ would be essential. The equation of motion of $\delta\psi$ becomes a constraint equation for $B_\mu$, i.e. $\partial_{\mu}\left(\sqrt{-g}\, \phi^2 B^{\mu} \right)=0$.

In both transverse and longitudinal channels, the equations for the fluctuations are PDEs. In this paper, however, we do not solve these PDEs directly. Instead, we try a different route. Exploiting the symmetry of the background, we rewrite each fluctuation by making use of a Fourier series expansion. In the basis of Fourier modes, the PDEs decompose into an infinite set of coupled ODEs. Because we work in the AdS-Schwarzschid black hole background, there is an advantage in doing so: Fourier modes with high momentum are exponentially suppressed in the bulk. For this reason, we expect that the behavior of the conductivity would be determined mostly by the couplings among relevant light modes.

\subsection{Transverse conductivity}
\label{sec:trancond}

Let us start from the transverse channel.  The equation of motion of $a_y(r,x)$ is
\begin{align}
\left( \partial_r^2 + \frac{f'}{f} \partial_r + \frac{1}{f^2}\partial_x^2 + \left( \frac{\omega^2}{f^2} - \frac{{\hphi}^2}{r^2 f} \right) \right) a_y &=0.
\label{trancond_eom_rx}
\end{align}
The novelty is the spatially dependent mass term proportional to ${\hphi}(r,x)^2$. 
Thanks to the symmetry of our background \eqref{phibg_cos_ansatz} and the assumption that the boundary electric field is homogeneous, the consistent Fourier decomposition of $a_y (r,x)$ is given by the following cosine series,
\begin{align}
a_y (r,x) &= a_y^{(0)}(r) + a_y^{(2)}(r) \cos (2Qx) + a_y^{(4)}(r) \cos (4Qx) + \cdots.
\label{trancond_cos_ansatz}
\end{align}
Plugging the ansatz \eqref{trancond_cos_ansatz} into \eqref{trancond_eom_rx} gives the following infinite set of coupled ODEs,
\begin{align}
 \left(  \partial_r^2  +\frac{f'}{f}\partial_r + \left( \frac{\omega^2}{f^2}- \frac{ \bphi^2 }{ 2r^2 f} \right) \right) a_y^{(0)} - \frac{ \bphi^2 }{4r^2 f} a_y^{(2)} &=0,  \label{ODE_transverse_1} \\
 \left(  \partial_r^2  +\frac{f'}{f}\partial_r + \left( \frac{\omega^2}{f^2}- \frac{4n^2Q^2}{f } - \frac{ \bphi^2 }{2 r^2 f} \right) \right)a_y^{(2n)} -\frac{ \bphi^2 }{4 r^2 f} \left( c_{2n-2} a_y^{(2n-2)}+a_y^{(2n+2)} \right) &= 0,
 \label{ODE_transverse}
\end{align}
where $n\ge 1$ and $c_{n}=\delta_{n,0}+1$. Through the mass term ${\hphi}(r,x)^2 a_y(r,x)$ in \eqref{trancond_eom_rx}, the striped background introduces couplings between different modes. As shown in \eqref{ODE_transverse_1} and \eqref{ODE_transverse}, they interact with a specific patter: $a_y^{(2n)}$ directly couples only to $a_y^{(2n\pm2)}$ whereas $a_y^{(0)}$ only couples to $a_y^{(2)}$ and not to $a_y^{(2l)}$ with $l>2$.

The electric field and the current at the boundary is obtained from the asymptotic behavior of $a_y(r,x)$.  In general, the series expansion of each mode at $r=0$ is
\begin{align}
a_y^{(2n)}(r) = a_{y,0}^{(2n)} + a_{y,1}^{(2n)} r + \cdots,
\end{align}
where $a_{y,0}^{(2n)}$ and $a_{y,1}^{(2n)}$ are the integration constants, and they correspond to the source and the vacuum expectation value of the boundary dual operator of $a_y^{(2n)}$. Sourcing a homogeneous boundary electric field implies that $a_{y,0}^{(0)}=E/i\omega$ and $a_{y,0}^{(2n)}=0$ for $n\ge1$, where $E$ is the magnitude of the electric field. These boundary conditions determine one integration constant for each field. To solve the second order Cauchy problem, we also impose the ingoing wave condition at the horizon, $ a_y^{(2n)}\propto f^{-i\omega r_h/3}$. The series expansion of $a_y^{(2n)}$ at the horizon is then computed. Finally, the ODEs are linear and therefore we can fix the overall normalization by choosing the value of one integration constant. We may set the magnitude of the electric field to one.

The induced current ${J}^y(x) = \sum_{n=0}^{\infty} a^{(2n)}_{y,1} \cos(2nQx)$ is a function of $x$, and so is the conductivity. We focus on the average of the conductivity in the $x$-direction, ${\hsigma}_T$. This is given by
\begin{align}
{\hsigma}_T(\omega) = - \frac{i}{\omega} \frac{ a_{y,1}^{(0)} }{a_{y,0}^{(0)} }\ .
\label{sigmaT_def}
\end{align}

\subsection{Longitudinal conductivity}
\label{sec:longcond}

Our main interest is on the conductivity in the direction longitudinal to the stripe. In particular, we focus on how it is affected by the interactions between gauge field and scalar fluctuations.
By considering the ansatz \eqref{longitudinal_fluctuation_ansatz}, the second order equations of motion become
\begin{align}
\left( \partial_r^2 + \frac{f'}{f} \partial_r + \left( \frac{ \omega^2}{f^2} - \frac{{\hphi}^2}{r^2 f} \right) \right) a_x - \frac{i\omega}{f^2} \partial_x a_t + \frac{1}{r^2 f} \left({\hphi} \partial_x - (\partial_x {\hphi}) \right) \psi &=0, \nonumber \\
\left( \partial_r^2 + \frac{1}{f}\partial_x^2 - \frac{{\hphi}^2}{r^2f} \right) a_t + \frac{i\omega}{f} \partial_x a_x - \frac{i\omega}{r^2 f}  {\hphi} \psi &=0, \nonumber \\
\left( \partial_r^2 + \left(\frac{f'}{f} -\frac{2}{r}\right) \partial_r + \frac{1}{f^2}\partial_x^2 +\left(   \frac{\omega^2}{f^2} - \frac{m^2 }{r^2 f} \right) \right) \psi - \frac{1}{f} \left({\hphi} \partial_x + 2(\partial_x {\hphi}) \right) a_x - \frac{i\omega}{f^2} {\hphi} a_t &=0\ .
\label{longcond_eom_rx}
\end{align}
In addition, there is a first order constraint coming from the $r$-component of the Maxwell equations,
\begin{align}
\partial_r \partial_x a_x + \frac{i \omega}{f} \partial_r a_t + \frac{1}{r^2} \left( (\partial_r {\hphi}) - {\hphi} \partial_r \right) \psi =0\ .
\label{Const_Long}
\end{align}
The derivative of the constraint with respect to $r$ is a linear combination of the three second order equations \eqref{longcond_eom_rx} upon using the background equation of motion \eqref{phi_r_eom}.

When we apply a homogeneous boundary electric field along the $x$ direction, the following Fourier decomposition ansatz can be used,
\begin{alignat}{2}
a_x (r,x) &= a_x^{(0)}(r) +{}&{} a_x^{(2)}(r) \cos (2Qx) +{}&{} a_x^{(4)}(r) \cos (4Qx) + \cdots , \nonumber \\
a_t (r,x) &=  & a_t^{(2)}(r) \sin (2Qx) +{}&{} a_t^{(4)}(r) \sin (4Qx) + \cdots , \nonumber \\
\psi (r,x) &=  & \psi^{(1)}(r) \sin (Qx) +{}&{} \psi^{(3)}(r) \sin (3Qx) + \cdots .
\label{longcond_cos_ansatz}
\end{alignat}
The general structure of the second order coupled ODEs is as follows,
\begin{alignat}{2}
D^{x} a_x^{(n)}\ & =\ \mathcal{M}^n_{\ m} a_x^{(m)} \, +\, Q\ \mathbb{A}^{ n}_{\ m} \psi^{(m)}\  +\,  &  i\frac{\omega Q }{f^2}\, n a_t^{(n)} \ , \nonumber \\
D^{t} a_t^{(n)}\ & =\ \mathcal{M}^n_{\ m} a_t^{(m)} \, -\, i \omega\ \mathbb{B}^{n}_{\ m} \psi^{(m)}\  +\, &  i\frac{\omega Q}{f}\, n  a_x^{(n)} \  ,\nonumber \\
D\, \psi^{(n)}\ & = \ Q \mathrm{A}^{ n}_{\ m}  a_x^{(m)} -\, i\omega\  \mathrm{B}^{n}_{\ m} a_t^{(m)}\ , & \label{Eqscheme}
\end{alignat} 
where $D^{x}, D^{t}, D$ are differential operators, and $\mathcal{M}, \, \mathrm{A}, \, \mathrm{B}, \, \mathbb{A},$ and $\mathbb{B}$ are $\bphi$-dependent non-diagonal matrices. For details, we refer the reader to Appendix \ref{sec:appeA}.

Interactions among different Fourier modes follow a specific pattern, which can be understood diagrammatically.
In order to visualize it, we associate a line to each non zero entry of the matrices appearing in the r.h.s of \eqref{Eqscheme}. This line relates two fields that are directly coupled. The precise form of the matrix entries is not important for the purpose of looking at the general structure of the interactions. As a result, we obtain the following diagram,
\begin{align}
\xymatrix{
\psi^{(1)}  \ar@{=}[r]|Q                            &  a_x^{(0)}  \ar@{~}[d]  &   & \psi^{(1)} \ar@{.}[dl]|\omega \\
\psi^{(3)}  \ar@{=}[r]|Q  \ar@{=}[dr]|Q  & a_x^{(2)} \ar@{-}[r]^{\omega Q}  \ar@{~}[d]   & a_t^{(2)} \ar@{~}[d]  \ar@{.}[r]|\omega  & \psi^{(3)} \ar@{.}[dl] |\omega \\
\psi^{(5)} \ar@{=}[r]|Q   \ar@{=}[dr]|Q  & a_x^{(4)}  \ar@{-}[r]^{\omega Q} \ar@{~}[d]   & a_t^{(4)} \ar@{~}[d]  \ar@{.}[r]|\omega &  \psi^{(5)} \ar@{.}[dl] |\omega \\
\vdots & \vdots & \vdots & \vdots \\
 },
\label{Diagram}
\end{align}
where the wavy, double, and dotted lines come from the mass matrix $\mathcal{M}$, the matrices $\mathrm{A}, \, \mathbb{A}$, and the matrices $\mathrm{B}, \, \mathbb{B}$, respectively. 
The diagram makes it evident that $\psi^{(1)}$ is not directly coupled to $a_x^{(2)}$, and therefore the block $\{ a_x^{(0)}, \psi^{(1)} \}$ can be singled out from the full pattern of interactions. 
Our main focus in this paper is to understand whether the coupling between $a_x^{(0)}$ and $\psi^{(1)}$ captures the most essential features of conductivity in the longitudinal channel.
We notice that generically the fields $\psi^{(2k+1)}$ for any $k\ge 1$ interact both with $a_x^{(2k)}$ and $a_x^{(2k+2)}$. Thus, the truncation to a greater number of fields is obtained after taking into account which fields, among $a_x^{(2k)}$, $a_t^{(2k)}$ and $\psi^{(2k+1)}$, are of the same order. In this case, also the constraint equation \eqref{Const_Long} has to be truncated consistently. 

Let's study when the truncation to the $\{ a_x^{(0)}, \psi^{(1)} \}$ block is regarded as a good approximation. The equations of $a_x^{(0)}$ and $\psi^{(1)}$ are given by
\begin{align}
\left(\partial_r^2 + \frac{f'}{f} \partial_r  - \left( \frac{ \bphi^2 }{ 2 r^2 f } - \frac{\omega^2}{f^2}  \right)\right) a_x^{(0)} &
= - \frac{Q }{f} \frac{\bphi}{r^2}\, \psi^{(1)}  +\frac{1}{4f} \frac{\bphi^2}{r^2}\, a_x^{(2)}\ , \label{Eqax0}\\
 \left( \partial_r^2+ \left( \frac{f'}{f}  -\frac{2}{r} \right) \partial_r- \left( \frac{m}{r^2 f} + \frac{Q^2}{f } -\frac{\omega^2}{f^2}  \right) \right) \psi^{(1)} &=  
 - 2\frac{Q\bphi}{ f }\, a^{(0)}_x  +  \frac{i \omega \bphi}{2 f^2}\, a^{(2)}_t \ .
 \label{Eqpsi1}
\end{align}
The truncation to the  $\{ a_x^{(0)}, \psi^{(1)} \}$ block would be valid when the interactions between $a_x^{(0)}$ and $\psi^{(1)}$ dominates over their interactions with $\{ a_x^{(2)}, a_t^{(2)} \}$. We first consider eq.~(\ref{Eqax0}). The coupling between $a_x^{(2)}$ and $a_x^{(0)}$ is proportional to $\bphi^2/r^2$ whereas the coupling between the rescaled field $\psi^{(1)}/r$ and $a_x^{(0)}$ is controlled by $Q \,\bphi/r$. Since $\bphi/r$ is a decreasing function of the radial coordinate, the ratio between these two coupings is bounded by its value at the boundary. 
This shows that the interaction between $a_x^{(0)}$ and $\psi^{(1)}$ dominates over that with $a_x^{(2)}$ if the condition $Q/V\gg 1$ is satisfied.  
The field $a_t^{(2)}$ is massive and not directly sourced by $a_x^{(0)}$. Hence, $a_t^{(2)}$ would be generically small compared with $a_x^{(0)}$.  Assuming that $a^{(2)}\propto f$ at the horizon, the decoupling of $a_t^{(2)}$ in eq.~\eqref{Eqpsi1} occurs in the small frequency limit, $\omega\ll Q$. 
Finally, the condition $Q/V \gg 1$ is trivially satisfied in the $Q\to \infty$ limit. In this regime, the modes $a_t^{(2n)}$ and $\psi^{(2n+1)}$ for $n\ge 1$ acquire a the large mass, and the system is somehow trivial. Also the bulk profile of the background scalar is exponentially suppressed in the $Q\to \infty$ limit

To solve the equations of motion, we need to fix boundary conditions.
The series expansion of each mode at $r \to 0$ is
\begin{align}
a_x^{(2n)}(r)  =  &\, a_{x,0}^{(2n)} + a_{x,1}^{(2n)} r + \cdots, \\
a_t^{(2n)}(r) = &\,a_{t,0}^{(2n)} + a_{t,1}^{(2n)} r+\cdots, \\
\psi^{(2n+1)}(r) =&\, \psi_{0}^{(2n+1)} r+ \psi_1^{(2n+1)} r^2+ \cdots \ ,
\end{align}
where $\psi^{(2n+1)}_{k}$ and $a^{(2n)}_{i,k}$ ($i=x,t$ and $k=0,1$) are the integration constants. The electric field at the boundary is
\begin{align}
f_{xt}=\lim_{r\rightarrow 0} \left( \partial_x a_t(r,x) -\partial_t a_x(r,x) \right).
\end{align}
A homogeneous boundary electric field is obtained by imposing the following set of constraints,
\begin{align}
a_{x,0}^{(0)}=\frac{E}{i\omega}\ ,\qquad
\ \Big( i\omega a_{x,0}^{(2n)} + 2n\, Q a_{t,0}^{(2n)} \Big) = 0\ , \quad \forall n\ge 1\ .
\end{align}
where $E$ is the magnitude of the electric field.
Since the system of ODEs is linear, we can fix the overall normalization by choosing the value of one integration constant. We may set $E=1$.
Fluctuations of the scalar field at the boundary are fixed by the standard or the alternative quantization. According to the choice of quantization, we demand that $\psi^{(m)}_0=0$ or $\psi^{(m)}_1=0$ for any $m$. At the horizon, we impose the ingoing wave condition with the perturbations behaving as,
\begin{alignat}{2}
a_x^{(2n)}=&\ \left(r-r_h\right)^{-i\omega r_h/3} \big(\, a^{(2n)}_{x+,\,0} &+ &\, a^{(2n)}_{x+,\,1}(r-r_h)+ \cdots \big)\ ,\nonumber \\
a_t^{(2n)}=&\ \left(r-r_h\right)^{-i\omega r_h/3}  \big(                             &\phantom{+}  &\, a^{(2n)}_{t+,\,1} (r-r_h)+\cdots \big)\ ,\nonumber \\
\psi^{(2n+1)}=&\ \left(r-r_h\right)^{-i\omega r_h/3} \big(\,  \psi^{(2n+1)}_{+,\,0}& + &\, \psi^{(2n)}_{+,\,1}(r-r_h)+ \cdots \big)\ . \label{hor_expa_Long}
\end{alignat}
The expansion above is determined by  the values of $a^{(2n)}_{x+,0}$ and $ \psi^{(2n+1)}_{+,0}$. The behavior of $a_t^{(2n)}$ is such that only the ingoing mode at the horizon is excited.

The average value of the conductivity along the direction of the stripe, ${\hsigma}_L(\omega) $, is obtained from the formula:
\begin{align}
{\hsigma}_L(\omega) = - \frac{i}{\omega} \frac{a_{x,1}^{(0)}}{a_{x,0}^{(0)}}\ .
\label{sigmaL_def}
\end{align}

\section{Analytic Calculations at Zero Temperature}
\label{sec:AdSCalculation}

In this section, we solve the system of coupled ODEs \eqref{Eqscheme} at zero temperature by considering a perturbative expansion in small $V/Q$, both in the transverse and the longitudinal channels. The point is that the curved background at zero temperature reduces to AdS, and the scalar background has a simple analytic form. This makes it possible to carry out the perturbative calculation analytically at each order.
Furthermore, at each order the perturbative calculation automatically implements a truncation to a finite set of Fourier modes, and heavy Fourier modes are suppressed by powers of $V/Q$.

The metric of the zero temperature AdS is given by \eqref{ads_sch_metric} with $f(r)=1$.
In the case that $m^2=-2$, the general solution of the equation \eqref{phi_r_eom} is a linear combination of $e^{\pm Q r}$.\footnote{  
When $m^2$ takes general values, the solution is given in terms of Bessel functions,
\begin{align}
\bphi(r) =  C_1 r^{3/2}\ J\left[\tilde{\Delta}, -i Q r \right]  + C_2 r^{3/2}\ Y\left[\tilde{\Delta}, -i Q r \right]  \nonumber
\end{align}
where $\tilde{\Delta}= \sqrt{m^2+9/4}$, and $J$ and $Y$ are Bessel functions. When $m^2=-2$, each Bessel function can be written as a linear combination of $e^{\pm Q r}$.}
Requiring that $\bphi(r)$ does not diverge exponentially as $r \to \infty$, we obtain
\begin{align}
\label{Sol_Zero_T}
\bphi(r)= V r e^{-Q r }.
\end{align} 
From the boundary expansion of this solution, we see that $\alpha=Q$. Thus, $\alpha=0$ in the homogeneous limit. This value can not be reached in the finite temperature probe limit, as seen in figure~\ref{fig:QalphaPlot}.

\subsection{Small lattice expansion in the transverse channel}
\label{sec:zeroT_small_Trans}

In the notation of section \ref{sec:trancond}, we can consistently expand each Fourier mode  $a_y^{(2n)}(r)$ in a power series in $V/Q$ by considering the ansatz:
\begin{alignat}{2}
a_y^{(0)}(r) &= a_y^{(0,0)}(r) \,+&\, a_y^{(0,2)}(r) \left(\frac{V}{Q}\right)^2 +&\, a_y^{(0,4)}(r) \left(\frac{V}{Q}\right)^4 + \cdots , \nonumber \\
a_y^{(2)}(r) &= & a_y^{(2,2)}(r) \left(\frac{V}{Q}\right)^2 +&\, a_y^{(2,4)}(r) \left(\frac{V}{Q}\right)^4 + \cdots , \nonumber \\
a_y^{(4)}(r) &= &&\, a_y^{(4,4)}(r) \left(\frac{V}{Q}\right)^4 + \cdots . \nonumber \\
\vdots \label{zeroT_pert_ansatz_trans}
\end{alignat}
The equations of these fields have the following form, 
\begin{align}
\partial_r^2 a_y^{(2n, i)}+(\omega^2-4n^2Q^2) a_y^{(2n,i)}=\mathcal{F}^{(2n,i)}(\omega,V,Q),
\label{pert_transv}
\end{align}
where $\mathcal{F}^{(0,0)}=0$, and $\mathcal{F}^{(2n,i)}$ are forcing terms. The field $a_y^{(0,0)}$ is a free field, and the fields $a_y^{(2n,i)}$ $(i>0)$ are solved iteratively from lower orders once the forcing terms are obtained. In general, the forcing term at level $(2n,i)$ is determined by a linear combination of lower order solutions $a_y^{(2m, j)}$ ($m\le n$ and $j\le i$). For example,
\begin{align}
\mathcal{F}^{(0,2)}=\mathcal{F}^{(2,2)}&= \frac{1}{2}e^{-2Q r}  a_y^{(0,0)} \ ,   \label{forc_y_2} \\
\mathcal{F}^{(0,4)}&=  \frac{1}{4}e^{-2Q r}\left( 2  a_y^{(0,2)} + a_x^{(2,2)} \right) \ .
\end{align}
The conductivity ${\hsigma}_T$ is defined from the asymptotic expansion of $a_y^{(0)}$, as shown in \eqref{sigmaT_def}. Given the ansatz \eqref{zeroT_pert_ansatz_trans}, ${\hsigma}_T$ is written as a perturbative series in $V/Q$, and its expansion is obtained after solving for the fields $a_y^{(0,i)}$ with $i\ge 0$.
The first few orders of this perturbative expansion are computed as follows: 
\begin{itemize}
\item
At the zeroth order, we need to solve for $a_y^{(0,0)}$. Its equation of motion, given by the $i=n=0$ case in \eqref{pert_transv}, reduces that of a free gauge field in flat space because of the $AdS_4$ background. The solution with the correct boundary conditions is an ingoing plane wave, 
\begin{align}
a_y^{(0,0)}(r)= A \, e^{i\omega r}
\label{Sol_AdS}
\end{align}  
where $A$ is an integration constant.
\item
At the second order, we solve for $a_y^{(0,2)}$. This field is sourced by  $a_y^{(0,0)}$. Plugging \eqref{Sol_AdS} in \eqref{forc_y_2}, we obtain its forcing term $\mathcal{F}^{(0,2)}$.
\item
At the fourth order, the forcing term $\mathcal{F}^{(0,4)}$ depends on $a_y^{(0,2)}$ and $a_y^{(2,2)}$. Hence, we first need to solve for $a_y^{(2,2)}$, whose forcing term is sourced by $a_y^{(0,0)}$. We use the solutions of $a_y^{(2,2)}$ and $a_y^{(0,2)}$ to determine $\mathcal{F}^{(0,4)}$, and we finally solve for $a_y^{(0,4)}$.
\end{itemize}
At higher orders, we repeat the iterative procedure following the strategy outlined above. Forcing terms are obtained step by step from the solutions at lower orders. How to fix the integration constants  at each order $i>0$, through boundary conditions, is explained in detail in appendix \ref{sec:appeB}.

The analytical solution of ${\hsigma}_T(\omega)$ at order $V^2/Q^2$ is  given as
\begin{align}
{\hsigma}_T(\omega) =  1+\frac{iQ^2}{4\omega (Q-i \omega) }\frac{V^2}{Q^2} + \mathcal{O}\left(\frac{V^4}{Q^4}\right).
\label{ZeroT_Cond_Trans}
\end{align}
The expression of the fourth order term is too cumbersome and is not shown here. Although the fourth order is the first place where couplings among $a_y^{(n,i)}$ with $n>0$ come in, they do not affect qualitative features of ${\hsigma}_T(\omega)$. The zeroth order contribution is the conductivity in AdS, obtained from the solution (\ref{Sol_AdS}): ${\hsigma}_{AdS}=1$. Our result \eqref{ZeroT_Cond_Trans} is qualitatively similar to the conductivity of an holographic superconductor close to the phase transition. The real part of ${\hsigma}_T(\omega)$ decreases at low frequencies with respect to ${\hsigma}_{AdS}$, and there is a delta function at $\omega=0$ with a positive weight.
This behavior is found because the bulk mass of $a_y$ is proportional to ${\hphi}(r,x)^2$ and has non zero average value of order $V^2$. This is enough to induce a superconducting current.

\subsection{Small lattice expansion in the longitudinal channel}
\label{sec:zeroT_small_Long}

In the notation of section \ref{sec:longcond}, the Fourier modes, $a_x^{(2n)}$, $a_t^{(2n)}$ and $\psi^{(2n+1)}$, are consistently expanded in powers of $V/Q$ as follows,
\begingroup
\allowdisplaybreaks
\begin{alignat}{2}
a_x^{(0)}(r) &= a_x^{(0,0)}(r) +{}&{} a_x^{(0,2)}(r) \left(\frac{V}{Q}\right)^2 +{}&{} a_x^{(0,4)}(r) \left(\frac{V}{Q}\right)^4 + \cdots , \nonumber \\
a_x^{(2)}(r) &= {}&{} a_x^{(2,2)}(r) \left(\frac{V}{Q}\right)^2 +{}&{} a_x^{(2,4)}(r) \left(\frac{V}{Q}\right)^4 + \cdots , \nonumber \\
a_x^{(4)}(r) &= {}&{}&{} a_x^{(4,4)}(r) \left(\frac{V}{Q}\right)^4 + \cdots , \nonumber \\
\vdots \nonumber \\
a_t^{(2)}(r) &= & a_t^{(2,2)}(r) \left(\frac{V}{Q}\right)^2 +{}&{} a_t^{(2,4)}(r) \left(\frac{V}{Q}\right)^4 + \cdots , \nonumber \\
a_t^{(4)}(r) &= &{}&{} a_t^{(4,4)}(r) \left(\frac{V}{Q}\right)^4 + \cdots , \nonumber \\
\vdots \nonumber \\
\psi^{(1)}(r) &= & \psi^{(1,1)}(r) \,\left(\frac{V}{Q}\right) \, +{}&{} \psi^{(1,3)}(r) \left(\frac{V}{Q}\right)^3 + \cdots , \nonumber \\
\psi^{(3)}(r) &= &{}&{} \psi^{(3,3)}(r) \left(\frac{V}{Q}\right)^3 + \cdots . \nonumber \\
\vdots \label{zeroT_pert_ansatz_long}
\end{alignat}%
\endgroup %
The second order equations of motion of these fields are schematically of the form,
\begin{align}
\left(\partial_r^2 +\omega^2\right) a_x^{(2n,i)}-\, i \omega Q\, 2n\, a_t^{(2n,i)}=&\ \mathcal{F}_{x}^{(2n,i)}(\omega,V,Q), \nonumber \\[4pt]
\left( \partial_r^2 - 4n^2 Q^2 \right) a_t^{(2n,i)}- i \omega Q\, 2n\, a_x^{(2n,i)}=&\ \mathcal{F}_t^{(2n,i)}(\omega,V,Q),\nonumber \\[4pt]
\left( \partial_r^2 - \frac{2}{r}\partial_r - \left(\frac{2}{r^2} +  (2n+1)^2 Q^2 -\omega^2\right)  \right) \psi^{(2n+1,k)}=&\ \mathcal{F}_\psi^{(2n+1,k)}(\omega,V,Q),
\label{Long_Systm}
\end{align} 
where $\mathcal{F}_{x}^{(2n,i)}$, $\mathcal{F}_{t}^{(2n,i)}$ and $\mathcal{F}_{\psi}^{(2n,i)}$ are the forcing terms in the longitudinal channel. 
Since $\mathcal{F}_x^{(0,0)}=0$, $a_x^{(0,0)}$ is a free field and its solution is the same as (\ref{Sol_AdS}), namely
\begin{align}
a_x^{(0,0)}(r)= A \, e^{i\omega r}.
\label{Sol_AdS_2}
\end{align}  
The solutions of all the other fields, $a_x^{(2n,i)}$, $a_t^{(2n,j)}$ and $\psi^{(2n+1,k)}$, are found iteratively as in the case of the transverse channel.  However, there is a significant difference from the transverse channel: The forcing terms are given as a linear combination of different fields at lower orders. For example,
\begin{align}
\mathcal{F}_\psi^{(1,1)}=&\ {2}Q r e^{-Q r} a_x^{(0,0)} \ ,    \\
\mathcal{F}_x^{(0,2)}=&\ \frac{1}{2}e^{-2Q r}  a_x^{(0,0)} - Q e^{-Qr}\frac{ \psi^{(1,1)} }{r}\ ,   \label{forcing_2} \\
\mathcal{F}_x^{(0,4)}=&\  \frac{1}{4}e^{-2Q r} \left(  2a_x^{(0,2)} + a_x^{(2,2)} \right) - Q e^{-Qr}\frac{ \psi^{(1,3)} }{r}\ .  \label{Fx_long}
\end{align}

The conductivity ${\hsigma}_L$, defined in \eqref{sigmaL_def}, is computed from the asymptotic expansion of the fields $a_x^{(0,i)}$ with $i\ge0$, and it is written as a perturbative series in powers of $V/Q$.
The zeroth order solution has been already found in \eqref{Sol_AdS_2}. Next few orders are solved as follows: 
\begin{itemize}
\item
At the second order, we solve for $a_x^{(0,2)}$. Its forcing term depends on $a_x^{(0,0)}$ and $\psi^{(1,1)}$. Therefore, we first need to solve for $\psi^{(1,1)}$. It is interesting to notice, from the expression of $\mathcal{F}_x^{(0,2)}$ given in \eqref{forcing_2}, that the conductivity at order $V^2$ already knows about the non trivial interaction between the scalar field and the gauge field. The first term in \eqref{forcing_2} comes from the mass matrix $\mathcal{M}$ and is analogous to the corresponding term in the transverse channel \eqref{forc_y_2}.
\item
At  the fourth order, the forcing term $\mathcal{F}_x^{(0,4)}$ depends on $a_x^{(0,2)}$, $a_x^{(2,2)}$ and $\psi^{(1,3)}$. In this case, we first solve for $a_x^{(2,2)}$ and $a_t^{(2,2)}$, whose forcing terms are sourced by $a_x^{(0,0)}$ and $\psi^{(1,1)}$, respectively. 
We then solve for $\psi^{(1,3)}$, whose forcing term is sourced by $a_x^{(0,2)}$ and $a_t^{(2,2)}$. 
Finally, we determine $\mathcal{F}_x^{(0,4)}$ and solve for $a_x^{(0,4)}$. As it is clear from this procedure, even though $\mathcal{F}_x^{(0,4)}$ does not depend directly on $a_t^{(2,2)}$, because $a_t^{(2,2)}$ couples to $a_x^{(2,2)}$, the solution of $a_x^{(0,4)}$ implicitly depends on $a_t^{(2,2)}$. 
\end{itemize}
We refer the reader to appendix \ref{sec:appeB} for further details about the calculation above and how to fix the integration constants at each order $i>0$. We just mention that,
according to the discussion in section \ref{sec:longcond}, the final result of the conductivity depends on the choice of quantization for the scalar field.

In the standard quantization we compute the conductivity up to the fourth order in $V/Q$. The result for $\omega<V$ is given by, 
\begin{align}
{\hsigma}_L(\omega)=&\ 1\,+\,  \frac{ i\omega + 2(Q+\sqrt{Q^2-\omega^2 } ) }{4(Q-i\omega) ( 2Q^2-\omega^2+ 2Q\sqrt{Q^2-\omega^2} ) } V^2\, +\,  
\mathcal{O}\left( \frac{V^4}{Q^4}\right) \label{cond_zeroT_Long_St}  \\ 
=&\ 1\,+\, \left(\frac{1}{4}+\frac{5i}{16}\frac{\omega}{Q} \, +\, 
\mathcal{O}(\omega^2)\right) \frac{V^2}{Q^2}+\,  \left(\frac{19}{512}-\frac{i}{128}\frac{Q}{\omega} \, +\, \mathcal{O}(\omega)\right) \frac{V^4}{Q^4} +\cdots,
 \label{cond_zeroT_Long_St1}
\end{align}
where the first line shows the exact expression at order $V^2/Q^2$, whereas the second line gives the expansion of ${\hsigma}_L(\omega)$ around $\omega=0$ including the fourth order corrections. 
Compared with the conductivity in transverse channel, there are two new features in \eqref{cond_zeroT_Long_St1}. The first is that the real part $\mathrm{Re}\, {\hsigma}_L$ is enhanced at small frequencies from the AdS value, and the second is that it also contains a delta function at $\omega=0$ with a negative weight, at order $V^4/Q^4$.\footnote{This statement follows from the standard Kramers-Kronig relation applied to the $i/\omega$ behavior in the imaginary part of ${\hsigma}(\omega)$.} At small frequencies $\mathrm{Re}\, {\hsigma}_L$ can be fitted with a Drude-like peak. However, it is important to stress that this Drude-like peak come together with the negatively weighted delta function at $\omega=0$, and therefore the dynamics of our system does not follow from an effective Drude theory. 

The appearance of the delta function can be understood from the point of view of spectral weight transfer. To illustrate better how this works, it is convenient to consider the conductivity in the alternative quantization, because a delta function with a negative weight appears already at order $V^2/Q^2$, and the calculation is simpler. In the alternative quantization, 
the result for $\hsigma_L(\omega)$, at frequencies $\omega < Q$, is
\begin{align}
{\hsigma}_L^-(\omega,Q) &= 1 + \sigma_L^{-(2)}(\omega,Q) + \mathcal{O}\left(\frac{V^4}{Q^4}\right) \nonumber \\
&= 1+ \left(  \frac{   P_1(\omega, Q) -i  P_2(\omega, Q)  }{ 4 (Q-i \omega) (Q^2-\omega^2 +Q  \sqrt{Q^2-\omega^2} )} \right)V^2 
+ \mathcal{O}\left(\frac{V^4}{Q^4}\right)   \label{cond_zeroT_Long1}\\
&=    1+\left( \frac{3}{4} -\frac{i}{\omega}\frac{Q}{2\omega} +\, \mathcal{O}(\omega) \right) \frac{V^2}{Q^2}\,+\, \mathcal{O}\left(\frac{V^4}{Q^4}\right) ,\label{cond_zeroT_Long_2}
\end{align}
where $\sigma_L^{-(2)}(\omega,Q)$ is the order $V^2/Q^2$ term, and
\begin{align}
P_1(\omega,Q)=&\ 4 Q -2\sqrt{Q^2-\omega^2}\ ,\\
P_2(\omega, Q)=&\ \frac{1}{\omega}(Q^2+3 \omega^2+3Q\sqrt{Q^2-\omega^2})\ .
\end{align}
As it is apparent from (\ref{cond_zeroT_Long_2}), there is a delta function in $\mathrm{Re}\, {\hsigma}^-_L\, $ at $\omega=0$ and its coefficient is negative.

The connection between the real part of the conductivity and the spectral weight density goes as follows. 
A Kubo type formula relates the conductivity ${\hsigma}(\omega)$ to the retarded Green's Function $G^R_{JJ}$ of the current. 
In this formalism, the real part of ${\hsigma}(\omega)$ is written as, 
\begin{align}
\mathrm{Re\, }{\hsigma}^{}(\omega)=
-\pi \ \mathrm{Re}\, G^R_{JJ}(0)\ \delta(\omega) +\frac{\mathrm{Im}\, G^R_{JJ}(\omega)}{\omega}\ ,
\label{real_part_kubo_formula}
\end{align}
where $\mathrm{Im}\, G^R_{JJ}$ is the spectral weight density. Then, the sum rule, 
\begin{align}
\int_0^\infty d\omega \, \mathrm{Re}( \sigma(\omega)-\sigma_\infty)=0, 
\label{SUMRULE}
\end{align}
says that the missing spectral weight at nonzero frequencies is balanced by the weight of a delta function at zero frequency. This is a global statements because the integral in \eqref{SUMRULE} is taken over all the frequencies.

It is a straight forward exercise to show that the sum rule \eqref{SUMRULE} is satisfied.
For example, we consider the alternative quantization case at order $V^2/Q^2$.
The result of the conductivity at all frequencies is
\begin{align}
{\hsigma}_L(\omega) =
\begin{cases}
1 + \sigma_L^{-(2)}(\omega,Q) + \mathcal{O}\left(V^4/Q^4\right) & (\omega<Q) \\
1 - \sigma_L^{-(2)}(i\omega,iQ) + \mathcal{O}\left(V^4/Q^4\right) & (\omega>Q)
\end{cases},
\label{alternative_order_VQ2_conductivity}
\end{align}
where $\sigma_L^{-(2)}(\omega,Q)$ is defined in \eqref{cond_zeroT_Long1}, and $\sigma_L^{-(2)}(i\omega,iQ)$ approaches zero as $\omega \to \infty$.
Using the real part of \eqref{alternative_order_VQ2_conductivity} with the delta function contribution at $\omega=0$ included, or using the RHS of \eqref{real_part_kubo_formula} computed directly from the field fluctuations, we find that the sum rule \eqref{SUMRULE} is satisfied.

The physical implication of this result becomes clear if we  consider the case of a system with a modulated superconducting state that has non zero average value of condensation. Such a system will have a constant value of the superfluid density, $n_s\neq 0$, and the conductivity ${\hsigma}(\omega)$ will be a function of $n_s$, $V$ and $Q$. Generically, $n_s$ and $\mathrm{Re}\,{\hsigma}$ will receive corrections due to lattice effects. According to our analysis, we would expect that the interaction between the charged finite momentum excitation and the current contributes to the final result both in the form of a Drude-like peak and as negative correction to $n_s$. In our model, $\delta\psi$ represents the charged low energy finite momentum excitation.
We mention that, in the case of a neutral holographic lattice, bulk phonons do not contribute to the shift of the spectral weight since $\mathrm{Im}\, {\hsigma}= \mathcal{O}(\omega)$. We also mention that the real part of the complex scalar field fluctuation, $\delta\eta$, is decoupled in our analysis. In the case of striped superconductivity, $\delta\eta$ is coupled to the gauge field fluctuations directly through the background charge density, and hence $\delta\eta$ may give nontrivial contributions to $n_s$. It would be interesting to single out the contributions of  $\delta\psi$ and $\delta\eta$ fluctuations in modulated superconductive systems, but this is beyond the scope of the present paper.

\section{Numerical Results at finite temperature}
\label{sec:numerical}

In this section, we compute the conductivity at finite temperature numerically. We consider a truncation to a minimal set of Fourier modes and solve the resulting coupled ODEs. The scalar background is parametrized in terms of the dimensionless quantities $T/V$ and $Q/V$. We consider a range of these parameters in which the truncation is a good approximation. For the transverse conductivity, we keep $a_y^{(0)}$ and $a_y^{(2)}$, and set $a_y^{(l)}$ with $l \ge 4$ to zero. For the longitudinal conductivity, we keep only $a_x^{(0)}$ and $\psi^{(1)}$.  We consider the scalar field to be dual to an operator of dimension $\Delta=2$, i.e. the standard quantization.

\subsection{Homogeneous conductivity}
\label{sec:homocond}

We briefly show the conductivity in the homogenous case, $\sigma_H$. The equation for the electromagnetic perturbation can be obtained from \eqref{trancond_eom_rx} by setting $Q=0$ and assuming the ansatz $a_y(r,x)=a(r)$. 
The ingoing wave condition is imposed at the horizon.

\begin{figure}[t]
\centering
\includegraphics[width=6cm]{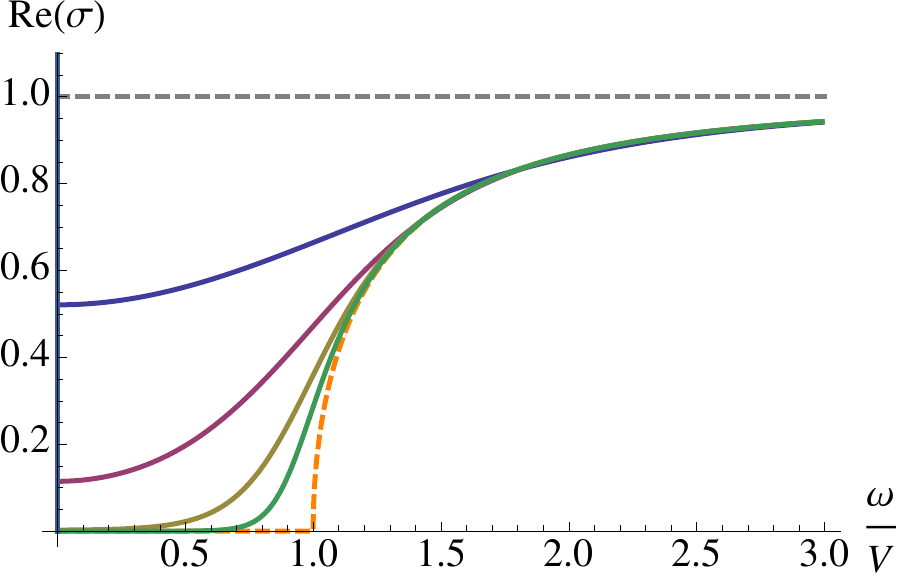}
\rule{.5cm}{0pt}
\includegraphics[width=6cm]{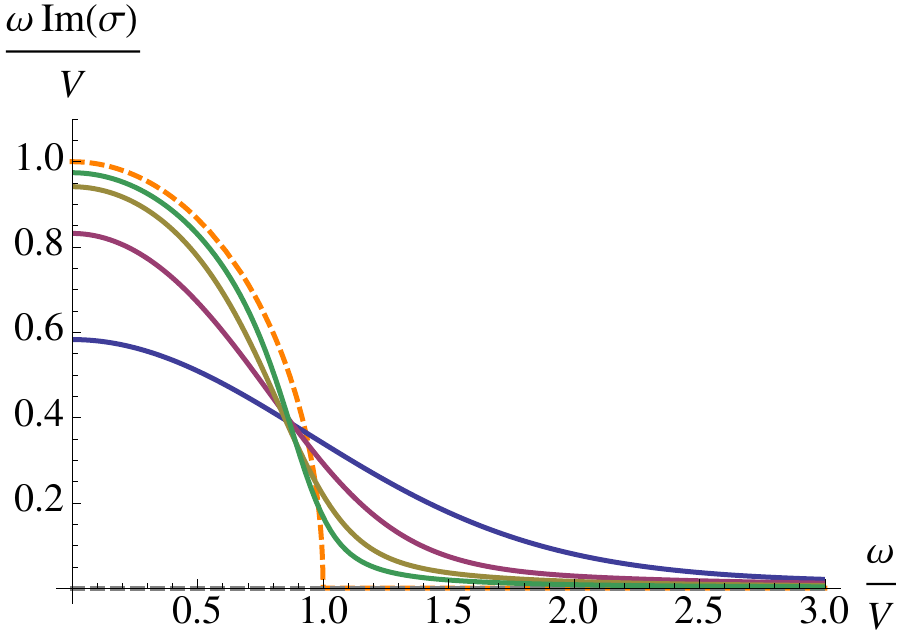}
\caption{Optical conductivity when $Q=0$. Left and right panels are the real and imaginary parts, respectively. The gray dashed lines are the result when $V = \langle \mathcal{O} \rangle=0$, the orange dashed lines are zero temperature result. In the left panel, from top to bottom, the solid lines correspond to $T/V = 1, \, 0.5, \, 0.25, \, 0.125$.} 
\label{fig:zeroQconductivity}
\end{figure}

Results for different values of $T/V$ are shown in figure~\ref{fig:zeroQconductivity}.
The real part of the conductivity becomes exponentially suppressed at small frequencies as $T/V\to 0$. At zero temperature, there is a hard gap of size $\omega/V=1$.
Thanks to the exact solution \eqref{Sol_Zero_T}, the zero temperature conductivity can be analytically obtained,\footnote{
This hard gap has been also investigated in \cite{Hartnoll:2008vx}.}
\begin{align}
\sigma(\omega) =
\begin{cases}
i\dfrac{\sqrt{V^2-\omega^2}}{\omega} & (\omega<V) \\[10pt]
\dfrac{\sqrt{\omega^2-V^2} }{\omega} & (\omega>V)
\end{cases},
\end{align}
and we also plot them.
For any value of $T/V$, the imaginary part of the conductivity diverges as $\mathrm{Im}(\sigma_H) \propto 1/\omega$ with a positive coefficient. Hence there is a delta function at $\omega=0$ in $\mathrm{Re}(\sigma_H)$.  As in the case of superconductivity, this delta function can be related to a superfluid current. In the limit $T/V \to \infty$, the conductivity approaches that of the 4D AdS-Schwarzschid black hole, $\sigma_{\mathit{Sch}}=1$.

\subsection{Transverse conductivity}
\label{sec:numtran}

\begin{figure}[t]
\centering
\includegraphics[width=6cm]{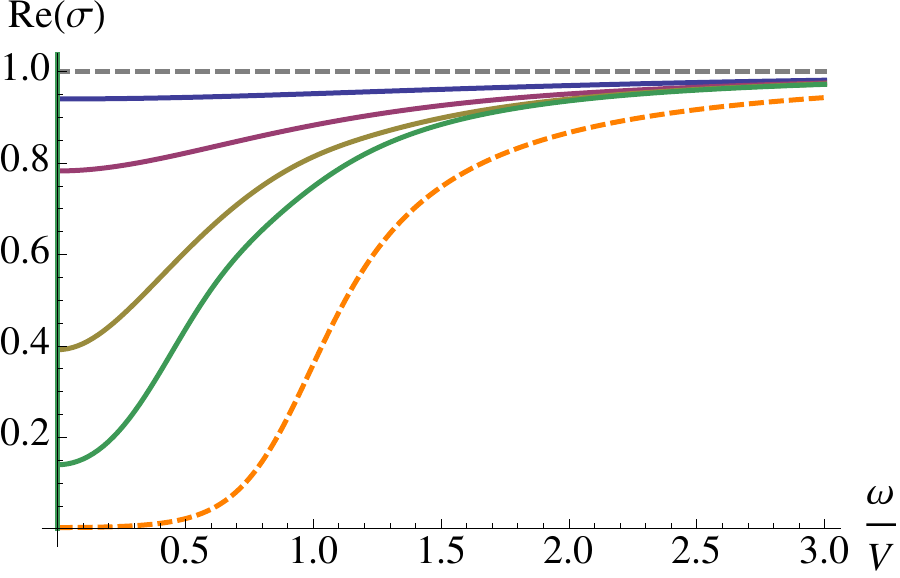}
\rule{.5cm}{0pt}
\includegraphics[width=6cm]{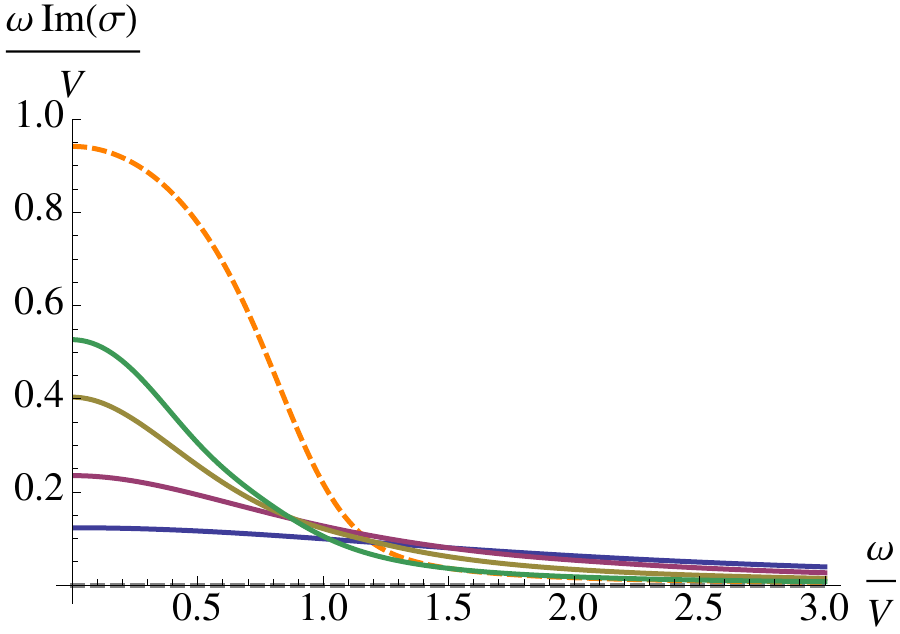}
\caption{The average of the optical conductivity in the transverse direction when $T/V=0.25$  and for various values of $Q/V$. Left and right panels are the real and imaginary parts of ${\hsigma}_T$, respectively. The gray dashed lines are the result when $Q/V =\infty$, the orange dashed lines are the result from the homogeneous case $Q=0$. In the left panel, from top to bottom the curves correspond to $Q/V = 2, \, 1, \, 0.5, \, 0.25$.} \label{fig:cyP4}
\end{figure}

We describe the behavior of the transverse conductivity ${\hsigma}_T(\omega)$ when $T/V$ is fixed and $Q/V$ is varied. In figure~\ref{fig:cyP4}, we show numerical results at $T/V=0.25$. When $Q/V\to\infty$, the conductivity approaches the constant value, $\sigma_{\mathit{Sch}}=1$.
When $Q=0$, the conductivity is given by the homogeneous result shown in figure~\ref{fig:zeroQconductivity}. For comparison, we also plot the results at $Q/V=0,\infty$ in figure~\ref{fig:cyP4}. As $Q/V$ decreases from the $Q/V=\infty$ limit, we can see a qualitative transition towards the $Q/V=0$ solution.  It is worth mentioning that when $Q/V$ is small, higher Fourier modes would be important to obtain quantitatively accurate results. 

Our results suggest that when $Q/V\gg 1$, the superconductivity is suppressed, whereas  in the opposite limit, $Q/V \ll 1$, the superfluid density increases with the stripe amplitude. This behavior can be qualitatively understood by extrapolating from the analytic zero temperature result of section \ref{sec:zeroT_small_Trans}. Because of the striped background, different modes within the Fourier decomposition of $a_y(x,r)$ are coupled, yet there is no effects like a Drude-like peak in figure~\ref{fig:cyP4}.  

\subsection{Longitudinal conductivity}
\label{sec:numlong}

\begin{figure}[t]
\centering
\includegraphics[width=6cm]{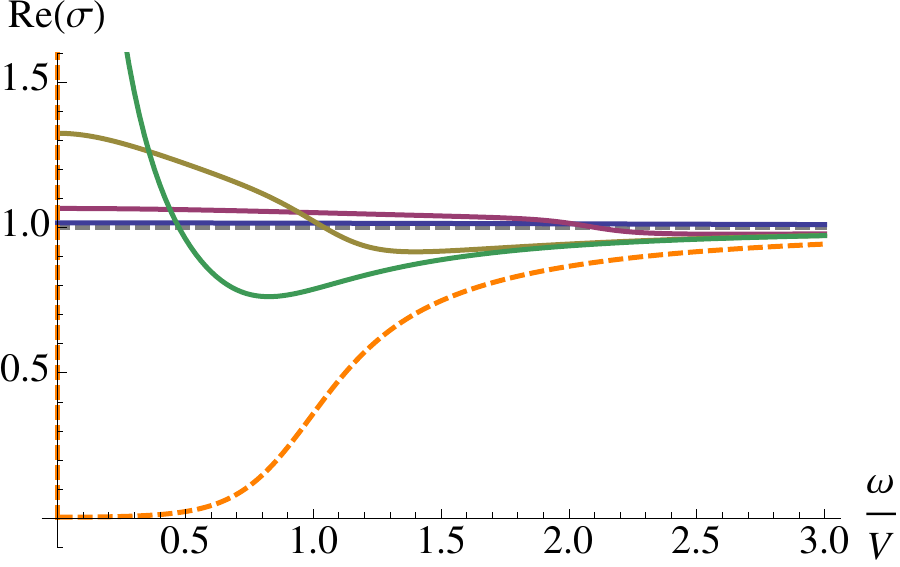}
\rule{.5cm}{0pt}
\includegraphics[width=6cm]{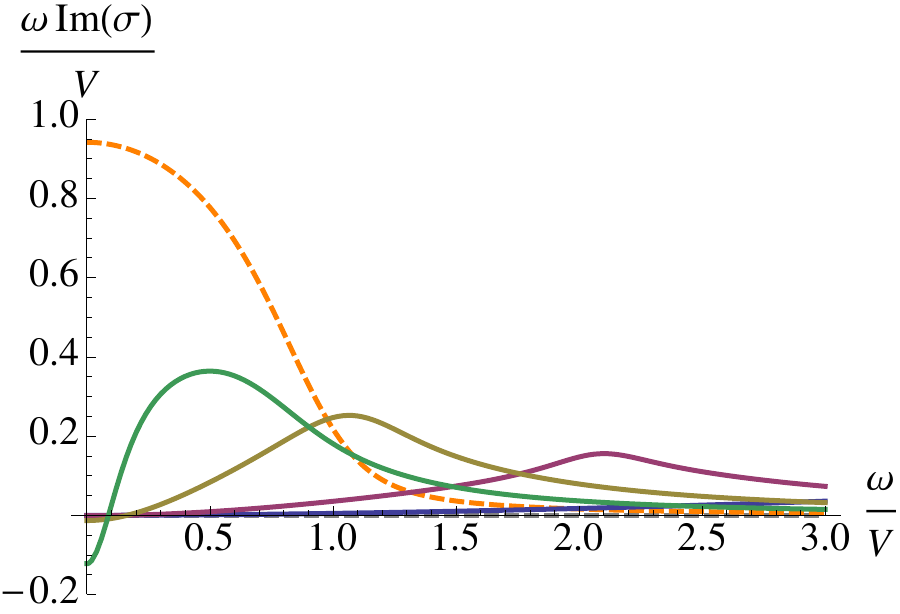}
\caption{The average of the optical conductivity in the longitudinal direction when $T/V=0.25$ and for various values of $Q/V$. Left and right panels are the real and imaginary parts of ${\hsigma}_L$, respectively. The gray dashed lines are the result when $Q/V =\infty$, the orange dashed lines are the result from the homogeneous case $Q=0$.  In the left panel,  from bottom to top the curves correspond to $Q/V = 4, \, 2, \, 1, \, 0.5$. For the value $Q/V=0.5$, $\mathrm{Re}\, \hsigma_L(0^+) = 4.05$ and lies outside the panel. }
\label{fig:cxP4}
\end{figure}

We now study the conductivity in the direction longitudinal to the stripe. In figure~\ref{fig:cxP4}, we show numerical results of ${\hsigma}_L(\omega)$ by varying $Q/V$ and keeping $T/V=0.25$. There are significant differences with respect to the transverse results, and it is now evident that the interaction between $a_x^{(0)}$ and $\psi^{(1)}$ is the most essential ingredient.

\begin{figure}[t]
\centering
\includegraphics[width=6cm]{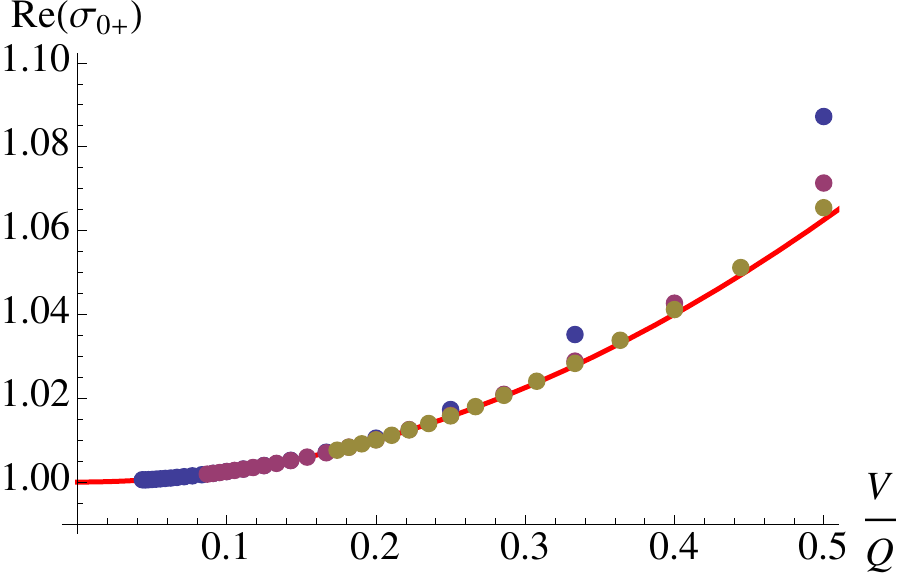}
\caption{$V/Q$ dependence of the $\omega \to 0$ limit of the real part of the average of the longitudinal conductivity. Dots are numerical results computed at several $Q/V$ and $T/V$: Blue, purple, and olive colors correspond to $T/V = 1, \, 0.5, \, 0.25$, respectively. The red line is $\mathrm{Re}(\sigma_{0+}) = 1+ (V/Q)^2/4$.}
\label{fig:VQDCPlot}
\end{figure}

Firstly, the real part of the conductivity is enhanced at small frequencies. This shape can be fitted with a Drude-peak, and the height of the peak increases as we go away from the $Q\to\infty$ limit. In the perturbative regime $V/Q\ll 1$, figure~\ref{fig:VQDCPlot} shows that, regardless of the temperature, the zero frequency limit of $\mathrm{Re}\, {\hsigma}_L$ behaves as,
\begin{align}
\left.\mathrm{Re}({\hsigma}_L)\right|_{\omega=0^+}\, = 1+\frac{1}{4}\frac{V^2}{Q^2}\ .
\label{VQDCfit}
\end{align}
Interestingly enough, this is the same behavior as we obtained in the zero temperature case.  We conclude that the zero frequency result (\ref{VQDCfit}) is temperature independent.

Secondly, the imaginary part of ${\hsigma}_L$ shows features in common with the zero temperature result. From the right panel of figure~\ref{fig:cxP4} we infer that $\mathrm{Re}\, {\hsigma}_L$ has a delta function with a negative weight. We also see that for each plot of $\, \omega\, \mathrm{Im}\,{\hsigma}_L$ there is a pronounced peak at finite $\omega/V$. In the large $Q$ limit the position of this peak is approximately $\omega_\mathit{peak} \simeq Q$, and it moves towards the origin as $Q/V$ is decreased. In the zero temperature computation of section~\ref{sec:zeroT_small_Long}, $\omega=Q$ is the locus where the first massive mode $\psi^{(1,1)}$ changes its behavior at the Poincare horizon: From being exponentially suppressed when $\omega<Q$, it becomes an ingoing wave when $\omega>Q$. At the same frequency $\omega_\mathit{peak}$, the real part of the conductivity crosses $\mathrm{Re}\, {\hsigma_L}=1$. As the peak moves towards the origin, the width of the Drude-like peak gets narrower. Eventually, in the $Q/V\to0$ limit, the Drude-like peak may become a delta function and the conductivity may coincide with the homogeneous result. To investigate the $Q\to0$ limit, however, it would be necessary to include higher order Fourier modes or to solve PDEs.\footnote{ In particular, an $N$-block truncation in the longitudinal channel is obtained by keeping the modes up to $\{ a_x^{(2N-2)}, a_t^{(2N-2)}, \psi^{(2N-1)} \}$ and setting other higher modes to zero. We have checked that a truncation to $N=2$ gives results in good agreement with figure~\ref{fig:cxP4}.}

\section{Conclusions and Discussion}
\label{sec:conclusion}

In this paper, we considered a simple holographic setup to study the optical conductivity in an inhomogeneous charged scalar background induced by a spatially varying source. In contrast to other constructions of holographic lattices, we set the charge density to zero. We applied a homogeneous electric field and computed the average value of the optical conductivity. Fluctuations of an inhomogeneous background generically
have finite momentum and may determine dissipative effects under certain circumstances. Identifying such circumstances in our model has been one focus of our paper. In particular, 
we paid special attention to the interaction between the complex scalar field and the gauge field fluctuations. The real part of the scalar fluctuation, $\delta\eta$, decoupled in our setup due to zero background charge density, and we focused on the dynamics of $\delta\psi$.

Our analysis relied on the use of a Fourier series expansion to reduce the equations of motion of the fluctuations, which were generically PDEs, to coupled ODEs. We then solved these ODEs instead of directly working with the PDEs. This approach turned out to be useful: In general, we could expect that perturbing an inhomogeneous background would lead to complicated nonlinear interactions among the fluctuations, and that we could understand the solution only by solving the PDEs. However, it might be the case that interactions within a subset of the modes in the Fourier decomposition of the fluctuations are sufficient to capture the dissipative dynamics of the inhomogeneous system qualitatively. We discussed that, in our model, the most essential interaction was the coupling between the charged scalar and the gauge field fluctuations coming from the nonzero $\partial_x$ derivative of the background. Indeed, the conductivity in the transverse channel turned out to be not dissipative, whereas the longitudinal channel showed interesting phenomena as follows.

We computed the optical conductivity in two cases. At zero temperature, we analytically computed the conductivity in a small amplitude expansion. In this case, we saw that higher Fourier modes were consistently suppressed, and we were able to compute the conductivity systematically from lower orders. At finite temperature, we numerically computed the conductivity by truncating the Fourier series to a relevant set of modes. 
In both cases, we found that a Drude-like peak and a delta function with a negative weight showed up in the longitudinal conductivity. By discussing the sum rule, we related these two features to a shift of low energy spectral weight.

In superconductors, the loss of phase coherence due to the presence of lattice impurities gives a negative correction to the superfluid density $n_s$. The negatively weighted delta function found in our analysis might also admit this interpretation since the $\delta\psi$ fluctuation actually drives a spatially-varying phase modulation of the induced condensate. Indeed, by applying a boundary electric field in the direction of the stripe, we saw that $\delta\psi$ is excited and backreact on the gauge current to determine a negative correction to the superfluid density $n_s$.  In particular, we explicitly saw in eqs.~\eqref{cond_zeroT_Long_St1} and \eqref{cond_zeroT_Long_2} that the order $V^2/Q^2$ contribution of $\delta\psi$ gave such a negative correction to the conductivity. Although our computation was done at zero charge density, we believe this statement to be generic.  

The Drude-like peak enhancement, instead, depends on the type of source deformation. In this paper we considered a scalar background with zero average value. As a generalization, we can consider a deformation $\phi_1(x) = V_0 + V \cos(Qx)$, in which $V_0$ induces a nonzero homogeneous source and gives a positive contribution to $n_s$. We made several numerical examinations in this case, and we observed that the Drude-like peak enhancement can be fine tuned and can disappear in the presence of nonzero $V_0$.

There would be many problems to be addressed in the future.
First, since it is difficult to use our Fourier decomposition approach to obtain quantitatively appropriate results at small $Q$, it would be necessary to solve PDEs in such a parameter region. Since the background scalar equation is linear in our model, it may be simple to construct more complicated inhomogeneous scalar backgrounds. For instance, we may consider a scalar configuration like a holographic Josephson Junction \cite{Horowitz:2011dz}.
It would be also important to include gravity backreaction. In our model, the zero temperature results in the presence of metric backreaction is expected to reduce to that given in this paper along the same line of \cite{Chesler:2013qla}. However, it would be possible to study different low temperature phases depending on the details of systems. 
For example, in the presence of an RG flow driven by a nontrivial scalar potential, one may try to construct a striped superconductor by using a double trace deformation or its generalization to inhomogeneous configurations. This would correspond to a phase with spontaneous translation and $U(1)$ symmetry breaking. Solving PDEs may not be avoided in this case. 
Last but not least, it would be also important to introduce nonzero charge density and reconsider the conductivity of striped superconductors where a $U(1)$ chemical potential is present \cite{,Hutasoit:2012ib}, by making use of a Fourier series expansion. This approach might be important for seeing what kind of interactions are singled out at finite chemical potential and understand what finally causes the power-law behavior seen in \cite{Horowitz:2012ky, Horowitz:2012gs,Horowitz:2013jaa}. Also it will clear up the comparison with the Q-lattices results of \cite{Donos:2013eha,Donos:2014uba}.

\acknowledgments
We would like to thank Dylan Albrecht, Mike Blake, Richard Davison, Aristomenis Donos, Matti J\"{a}rvinen, Giuseppe Policastro and David Tong for valuable discussions. We are grateful to DAMTP at the University of Cambridge and Leiden University for hospitality. 
The works of the authors are supported in part by European Union's Seventh Framework Programme under grant agreements ((FP7-REGPOT-2012-2013-1) no 316165, PIF-GA-2011-300984, the EU program ``Thales'' and ``HERAKLEITOS II'' ESF/NSRF 2007-2013 and is also co-financed by the European Union (European Social Fund, ESF) and Greek national funds through the Operational Program ``Education and Lifelong Learning'' of the National Strategic Reference Framework (NSRF) under ``Funding of proposals that have received a positive evaluation in the 3rd and 4th Call of ERC Grant Schemes''.

\appendix

\section{Fourier Decomposition in the Longitudinal Channel}
\label{sec:appeA}
In this appendix, we give the details of eq.~\eqref{Eqscheme}.
The second order ODEs for the modes $\{ a_x^{(n)}, a_t^{(n)}, \psi^{(m)} \}$ are, 
\begin{align}
D\, \psi^{(m)}-  \frac{i\omega \bphi}{2f^2} \left( a^{(m+1)}_t + a^{(m-1)}_t  \right) + \frac{Q\bphi}{2f} \left( (m+1) c_{m-1} a^{(m-1)}_x +  (m-1) a^{(m+1)}_x \right) =& 0\ ,
\nonumber\\
D^x a_x^{(n)}  - \frac{\bphi^2}{4 r^2 f} \left( c_{n-2} a^{(n-2)}_x + a^{(n+2)}_x  \right) - \frac{i \omega Q}{f^2} n a^{(n)}_t  +  \frac{Q \bphi }{2 r^2 f} \left( (n+2) \psi^{(n+1)} + (n-2) \psi^{(n-1)}  \right) =& 0\ ,
\nonumber\\
D^t a_t^{(n)} - \frac{\bphi^2}{4 r^2 f } \left( a^{(n-2)}_t + a^{(n+2)}_t  \right) - \frac{i\omega Q}{f } n a^{(n)}_x  - \frac{i w  \bphi}{2 r^2 f } \left(  \psi^{(n+1)} +  \psi^{(n-1)} \right) = & 0\ ,
\end{align}
where $m=2k-1$ and $n=2k$ with $k\ge 1$, $c_{n}=\delta_{n,0}+1$, $a_t^{(0)}=0$ and we have defined the following differential operators,
\begin{eqnarray}
D^t a_t^{(n)}  & = & \partial_r^2 a_t^{(n)}  - \left( \frac{ \bphi^2 }{ 2 r^2 f } + \frac{n^2Q^2}{f }  \right) a_t^{(n)}\ ,
\nonumber\\
D^x a_x^{(n)} & = &  \partial_r^2 a_x^{(n)} + \frac{f'}{f} \partial_r a_x^{(n)} - \left( \frac{ \bphi^2 }{ 2 r^2 f } - \frac{\omega^2}{f^2}  \right) a_x^{(n)}\ ,
\nonumber\\
D\, \psi^{(n)} & = & \partial_r^2 \psi^{(n)} + \left( \frac{f'}{f}  -\frac{2}{r} \right) \partial_r \psi^{(n)} - \left( \frac{m}{r^2 f} + \frac{n^2Q^2}{f } -\frac{\omega^2}{f^2}  \right) \psi^{(n)}\ . 
\end{eqnarray}
The constraint (\ref{Const_Long}) gives the following relation valid for any $n\ge 1$,
\begin{align}
-i\frac{\omega}{f} \partial_r a_t^{(2n)}\, +\, 2n Q \partial_r a_x^{(2n)} +\frac{1}{2r^2}(\bphi\partial_r-\bphi')\left( \psi^{(2n-1)}+\psi^{(2n+1)} \right)=0\ .
\label{Constr_mode}
\end{align}
The $a_x^{(0)}$ mode does not appear in (\ref{Constr_mode}).  It is important to notice that the $N$-block truncation obtained by setting to zero all the modes $\{ a_x^{(2k)}, a_t^{(2k)}, \psi^{(2k+1)} \}$ for $k\ge N > 1$ is consistent with the constraint equation of $a_t^{(2N-2)}$, $a_x^{(2N-2)}$ and $\psi^{(2N-2\pm 1)}$.

\section{Perturbative Expansion at Zero Temperature}
\label{sec:appeB}

In section \ref{sec:AdSCalculation} we solved the system of coupled ODEs, both in the transverse and the longitudinal channel, 
considering a perturbative expansion of the fields in the parameter $V/Q$.  In this appendix we explain how to fix the boundary conditions at each order in the perturbative computations. 
\paragraph{The transverse channel.}
The equation of the field $a_y^{(2n, i)}$ is,   
\begin{align}
\label{equa_appb_1}
\partial_r^2 a_y^{(2n, i)}+(\omega^2-4n^2Q^2) a_y^{(2n,i)}=\mathcal{F}^{(2n,i)}(\omega,V,Q)\ ,
\end{align}
where the l.h.s is given by the second order linear operator $L_n:= \partial_r^2+(\omega^2-4n^2Q^2)$ acting on $a_y^{(2n, i)}$, whereas the r.h.s is given by a forcing term whose specific form depends on the fields $a_y^{(2m, j)}$ at lower orders. The most general solution of (\ref{equa_appb_1}) is a linear combination of the two independent kernel solutions, $L_n f^{(2n,i)}=0$, plus a particular solution which is forced by $\mathcal{F}^{(2n,i)}$.
 Let's consider the case of $a_x^{(0,2)}$ for concreteness. The forcing term is $\mathcal{F}^{(0,2)}=1/2 e^{-2Q r} a_y^{(0,0)}$ with $a_y^{(0,0)}= A e^{i\omega r}$, as given in \eqref{Sol_AdS}, and the most general solution is 
\begin{align}
a_y^{(0,2)}= C^{(0,2)}_1 e^{-i\omega r}+C^{(0,2)}_2 e^{i\omega r}+ \frac{E}{8} \frac{e^{-2Q r+i\omega r}}{Q-i\omega}\ ,
\end{align}
where the first two terms are the kernel of $L_0$, i.e. solutions of $L_0 f=0$. 
The integration constant $C^{(0,2)}_1$ is set to zero by imposing the ingoing wave boundary condition. To fix the solution completely we need to specify $C^{(0,2)}_2$. 
In general, the solutions of $L_n f^{(2n,i)}=0$ are 
\begin{align}
 f^{(2n,i)}= \begin{cases}
 C^{(2n,i,-)}_1 e^{-r \sqrt{ 4n^2Q^2-\omega^2} } + C^{(2n,i,-)}_2 e^{ r \sqrt{ 4n^2Q^2-\omega^2} } & \mathrm{for\ }\omega<2nQ\ ,\\
 C^{(2n,i,+)}_1 e^{-i r \sqrt{\omega^2- 4n^2Q^2} } + C^{(2n,i,+)}_2 e^{ i r \sqrt{\omega^2- 4n^2Q^2} } & \mathrm{for\ } \omega>2nQ\ .
\end{cases}
\label{appb_kernel_trans}
\end{align}
When $\omega> 2nQ$, the ingoing wave condition is imposed, and this sets $C^{(2n,i,+)}_1=0$. For $\omega< 2 n Q$, we require the field not to diverge exponentially as $r\to \infty$ and therefore we set $C^{(2n,i,-)}_2=0$. In all these cases, we are left with one integration constant to specify for each field $a_y^{(2n,i)}$. To completely fix the solutions we require the boundary electric field to be homogeneous, 
\begin{align}
\label{appB_cond}
a_y^{(0,0)}(0)=A= \frac{E}{i\omega}, \qquad\quad a_y^{(2n,i)}(0)=0\qquad \forall n\ge 0,\, i>0\ .
\end{align} 
In this way, we allow the current $J^y$ to be corrected order by order and we guarantee that the boundary electric field is homogeneous, i.e.
\begin{align}
a_y(r)=\frac{E}{i\omega}+ J^y\left(\frac{\omega}{Q}, \frac{V}{Q}\right) r + O(r^2)\ .
\end{align} 

\paragraph{The longitudinal channel.}
The fields we need to solve for are $a_x^{(2n,i)}$, $a_t^{(2n,j)}$ and $\psi^{(2n+1,k)}$. It is convenient to first discuss the case of $\psi^{(2n+1,i)}$, whose general equation can be written in the following way, 
\begin{align}
L^\psi_{\ n}\psi^{(2n+1,i)}=&\ \mathcal{F}^{(2n,i)}(\omega,V,Q)\ ,\\
L^\psi_{\ n}:=&\ \left( \partial_r^2 - \frac{2}{r}\partial_r - \left(\frac{2}{r^2} +  (2n+1)^2 Q^2 -\omega^2\right) \right)  ,
\end{align}
The kernel of $L_{\ n}^\psi$ is obtained from \eqref{appb_kernel_trans} by noticing that $L_{\ n}^\psi (r g) \propto L_n (g) $. The solution of $L_{\ n}^\psi f^{(2n+1,i)}=0$, with the correct boundary conditions at $r\to \infty$, is then, 
\begin{align}
f^{(2n+1,i)}(r) =
\begin{cases}
C_\psi^{(2n+1,i,-)} r e^{- r\sqrt{(2n+1)^2Q^2 - \omega^2}} &  \mathrm{for\ }\omega<(2n+1)Q\ , \\
C_\psi^{(2n+1,i,+)} r e^{i r \sqrt{\omega^2 - Q^2(2n+1)^2 }} &  \mathrm{for\ } \omega>(2n+1)Q\ .
\end{cases}
\end{align}
The integration constant $C_{\psi}$ is fixed by imposing standard or alternative quantization on the solution of $\psi^{(2n+1,i)}$.
Finding the solution of $a_x^{(0,i)}$ is also straightforward since the equation is determined only by the linear operator $L_0$ and the forcing terms, 
\begin{align}
\left( \partial_r^2 +\omega^2\right) a_x^{(0,i)}=\mathcal{F}_x^{(0,i)}(\omega,V,Q)\ .
\end{align}
The kernel solutions are $e^{\pm i\omega r}$ and the integration constants are fixed by imposing the ingoing wave condition as $r\to \infty$ and
\begin{align}
a_x^{(0,i)}(0)=0\qquad \forall\, i>0\ .
\end{align} 
Solving for the fields $a_x^{(2n)}$ and $a_t^{(2n)}$ with $n\ge 1$ is more involved because they are coupled,
\begin{align}
\left(\partial_r^2 +\omega^2\right) a_x^{(2n,i)}-\, i \omega Q\, 2n\, a_t^{(2n,i)}=&\ \mathcal{F}_x^{(2n,i)}(\omega,V,Q), \label{appb_eq_ax} \\
\left( \partial_r^2 - 4n^2 Q^2 \right) a_t^{(2n,i)}- i \omega Q\, 2n\, a_x^{(2n,i)}=&\ \mathcal{F}_t^{(2n,i)}(\omega,V,Q). \label{appb_eq_at}
\end{align} 
The strategy we adopt is similar to \cite{Policastro:2002se}. 
In order to proceed, we consider the constraint equation in the perturbative expansion. This is given by,
\begin{align}
-i{\omega}\partial_r a_t^{(2n,i)}\, +\, 2n Q \partial_r a_x^{(2n,i)} =\mathcal{G}^{(2n,i)}(\omega,V,Q) \ .
\label{appb_long_constraint}
\end{align}
where $\mathcal{G}^{(2n,i)}$ explicitly depends on $\psi^{(2m-1,i)}$ with $m\le n$. 
Since the first derivatives of $a_x^{(2n)}$ and $a_t^{(2n)}$ are related by the constraint equation, we use this to write a single equation for $da_t^{(2n,i)} \equiv \partial_r a_t^{(2n,i)}$. 
Then, the equation of $da_t^{(2n,i)}$ has the form,
\begin{align}
\left( \partial_r^2+\omega^2 -4n^2 Q^2\right) da_t^{(2n.i)}=\partial_r \mathcal{F}^{(2n.i)} +i\omega \mathcal{G}^{(2n.i)}\ ,
\label{appb_daequation}
\end{align}
where on l.h.s we recognize the linear operator $L_n$. The kernel solution is given by \eqref{appb_kernel_trans}, and the boundary condition at $r\to \infty$ is such that the field does not diverge exponentially when $\omega<2nQ$ and is the ingoing wave when $\omega>2nQ$, as in the case of $a_y^{(2n,i)}$. This fixes one of the integral constants of \eqref{appb_daequation}. After solving for $da_t^{(2n,i)}$, we automatically obtain $\partial_r a_x^{(2n,i)}$ from \eqref{appb_long_constraint}, and by taking the derivatives we obtain $\partial_r^2 a_t^{(2n,i)} $and $\partial_r^2 a_x^{(2n,i)}$. The equations \eqref{appb_eq_ax}-\eqref{appb_eq_at} are now algebraic equations for $a_x^{(2n,i)}$ and $a_t^{(2n,i)}$. They are not independent and provide a solution for the linear combination $E_x^{(2n,i)}\equiv i\omega a_x^{(2n,i)}+ 2n Q a_{t}^{(2n,i)}$, which is the bulk electric field. To fix all integration constants in eqs.~\eqref{appb_eq_ax}-\eqref{appb_long_constraint}, we need two more conditions. We require that the boundary electric field is homogeneous,
\begin{align}
E_x^{(2n,i)}|_{r=0}=\Big( i\omega a_{x}^{(2n,i)} + 2n Q a_{t}^{(2n,i)} \left. \Big) \right|_{r=0} = 0\ , \quad \forall\, n\ge 1\quad i>0 \ .
\end{align}
We also set the chemical potential to zero by requiring (See  e.g.~Ref.~\cite{Nakamura:2007nk}.)
\begin{align}
\mu^{(2n,i)} \equiv \int_{\infty}^0 \mathrm{d}{r}\, da_t^{(2n,i)}=0\ .
\end{align}

\providecommand{\href}[2]{#2}\begingroup\raggedright\endgroup

\end{document}